\documentclass[%
reprint,
superscriptaddress,
amsmath,amssymb,
prc,
floatfix,]%
{revtex4-1}

\usepackage{CJK}
\usepackage{graphicx}
\usepackage{dcolumn}
\usepackage{bm}
\usepackage{enumerate}
\usepackage[bookmarks=true,colorlinks=true,linkcolor=blue,
           citecolor=blue,hypertex,breaklinks=true,urlcolor=blue]{hyperref}
\usepackage{booktabs}
\usepackage{array}
\begin{document}

\title{Nuclear charge radius predictions by kernel ridge regression with odd-even effects}

\begin{CJK*}{GBK}{}

 \author{Lu Tang}
 \affiliation{Mathematics and Physics Department,
              North China Electric Power University, Beijing 102206, China}

 \author{Zhen-Hua Zhang }
 \email{zhzhang@ncepu.edu.cn}
 \affiliation{Mathematics and Physics Department,
              North China Electric Power University, Beijing 102206, China}
 \affiliation{Hebei Key Laboratory of Physics and Energy Technology,
              North China Electric Power University, Baoding 071000, China}

\date{\today}

\begin{abstract}
The extended kernel ridge regression (EKRR) method with odd-even effects was adopted to improve the description of the nuclear
charge radius using five commonly used nuclear models.
These are: (i) the isospin dependent $A^{1/3}$ formula,
(ii) relativistic continuum Hartree-Bogoliubov (RCHB) theory,
(iii) Hartree-Fock-Bogoliubov (HFB) model HFB25,
(iv) the Weizs\"acker-Skyrme (WS) model WS$^\ast$, and
(v) HFB25$^\ast$ model.
In the last two models, the charge radii were calculated using a five-parameter formula
with the nuclear shell corrections and deformations obtained from the WS and HFB25 models, respectively.
For each model, the resultant root-mean-square deviation for the 1014 nuclei
with proton number $Z \geq 8$ can be significantly reduced
to 0.009-0.013~fm after considering the modification with the EKRR method.
The best among them was the RCHB model, with a root-mean-square deviation of 0.0092~fm.
The extrapolation abilities of the KRR and EKRR methods for the neutron-rich region were examined
and it was found that after considering the odd-even effects,
the extrapolation power was improved compared with that of the original KRR method.
The strong odd-even staggering of nuclear charge radii of Ca and Cu isotopes
and the abrupt kinks across the neutron $N=126$ and 82 shell closures were also
calculated and could be reproduced quite well by calculations using the EKRR method.
\end{abstract}

\maketitle
\end{CJK*}

\section{Introduction}\label{sec:intro}

The nuclear charge radius, similar to other quantities such as  the binding energy and half-life,
is one of the most basic properties reflecting the important characteristics of atomic nuclei.
Assuming a constant saturation density inside the nucleus,
the nuclear charge radius is usually described by the $A^{1/3}$ law, where $A$ is the mass number.
By studying  the charge radius, information on the
nuclear shells and subshell structures~\cite{Angeli2015_JPG42-055108, Gorges2019_PRL122-192502},
shape transitions~\cite{Wood1992_PR215-101, Cejnar2010_RMP82-2155},
the neutron skin and halos~\cite{Tanihata1985_PRL55-2676, Tanihata2013_PPNP68-215, Meng2015_JPG42-093101},
etc., can be obtained.

With improvements in the experimental techniques and measurement methods,
various approaches have been adopted for measuring
the nuclear charge radii~\cite{Cheal2010_JPG37-113101, Campbell2016_PPNP86-127}.
To date, more than 1000 nuclear charge radii have been measured~\cite{Angeli2013_ADNDT99-69, Li2021_ADNDT140-101440}.
Recently, the charge radii of several very exotic nuclei have
attracted interest, especially the strong odd-even staggering (OES) in some isotope chains
and the abrupt kinks across neutron shell closures~\cite{Goddard2013_PRL110-032503,
Hammen2018_PRL121-102501, Ruiz2016_NatPhys12-594, Miller2019_NatPhys15-432,
Gorges2019_PRL122-192502, Groote2020_NatPhys16-620, Goodacre2021_PRL126-032502,
Reponen2021_NC12-4596, Koszorus2021_NatPhys17-439, Malbrunot-Ettenauer2022_PRL128-022502,
Geldhof2022_PRL128-152501}, which provide a benchmark for nuclear models.

Theoretically, except for phenomenological formulae~\cite{Bohr1969_Book, Zeng1957_ActaPhysSin13-357,
Nerlo-Pomorska1993_ZPA344-359, Duflo1994_NPA576-29, Zhang2002_EPJA13-285,
Lei2009_CTP51-123, Wang2013_PRC88-011301, Bayram2013_APPB44-1791},
various methods, including
local-relationship-based models~\cite{Piekarewicz2010_EPJA46-379, Sun2014_PRC90-054318,
Bao2016_PRC94-064315, Sun2017_PRC95-014307, Bao2020_PRC102-014306, Ma2021_PRC104-014303},
macroscopic-microscopic models~\cite{Buchinger1994_PRC49-1402, Buchinger2001_PRC64-067303,
Buchinger2005_PRC72-057305, Iimura2008_PRC78-067301},
nonrelativistic ~\cite{Stoitsov2003_PRC68-054312, Goriely2009_PRL102-242501,
Goriely2010_PRC82-035804, Reinhard2017_PRC95-064328}
and relativistic mean-field model~\cite{Lalazissis1999_ADNDT71-1, Geng2005_PTP113-785, Zhao2010_PRC82-054319,
Xia2018_ADNDT121-122-1, Zhang2020_PRC102-024314, An2020_PRC102-024307,
Perera2021_PRC104-064313, Zhang2022_ADNDT144-101488, An2023_NST34-119}
were used to systematically investigate nuclear charge radii.
In addition, the $ab$-initio no-core shell model was adopted for
investigating this topic~\cite{Forssen2009_PRC79-021303, Choudhary2020_PRC102-044309}.
Each model provides fairly good descriptions of the nuclear charge radii across the nuclear chart.
However, with the exception of models based on local relationships,
all of these methods have root-mean-square (RMS) deviations larger than 0.02~fm.
It should be noted that few of these models can reproduce
strong OES and abrupt kinks across the neutron shell closure.
To understand these nuclear phenomena, a more accurate description of nuclear charge radii is required.

Recently, due to the development of high- performance computing,
machine learning methods have been widely adopted for investigating
various aspects of nuclear physics~\cite{Bedaque2021_EPJA57-100, Boehnlein2022_RMP94-031003,
He2023_NST34-88, He2023_SciChinaPAM66-282001, Gao2021_NST32-109}.
Several machine learning methods have been used to improve the description of nuclear
charge radii, such as artificial neural networks ~\cite{Akkoyun2013_JPG40-055106,
Wu2020_PRC102-054323, Shang2022_NST33-153, Yang2023_PRC108-034315},
Bayesian neural networks~\cite{Utama2016_JPG43-114002, Neufcourt2018_PRC98-034318,
Ma2020_PRC101-014304, Dong2022_PRC105-014308, Dong2023_PLB838-137726},
the radial basis function approach~\cite{Li2023_NPR_40-31},
the kernel ridge regression (KRR)~\cite{Ma2022_CPC46-074105}, etc.
By training a machine learning network using radius residuals,
that is, the deviations between the experimental and calculated nuclear charge radii,
machine learning methods can reduce the corresponding rms deviations to 0.01-0.02~fm.

The KRR method is one of the most popular machine-learning approaches, with the extension of
ridge regression for nonlinearity~\cite{Kim2012_IEEE42-1011, Wu2017_IEEE47-3916}.
It was improved by including odd-even effects and gradient kernel functions
and provided successful descriptions of various aspects of nuclear physics, such as of the
nuclear mass~\cite{Wu2020_PRC101-051301, Wu2021_PLB819-136387, Guo2022_Symmetry14-1078,
Wu2022_PLB834-137394, Du2023_ChinPhysC47-074108},
nuclear energy density functionals~\cite{Wu2022_PRC105-L031303},
and neutron-capture reaction cross-sections~\cite{Huang2022_CTP74-095302}.
In the present study, the extended KRR (EKRR) method with odd-even effects included
through remodulation of the KRR kernel function~\cite{Wu2021_PLB819-136387}
is used to improve the description of the nuclear charge radius.
Compared with the KRR method, the number of weight parameters did not increase in the EKRR method.

The remainder of this paper is organized as follows.
A brief introduction to the EKRR method is presented in Sec.~\ref{sec:ekrr}.
The numerical details of the study are presented in Sec.~\ref{sec:details}.
The results obtained using the KRR and EKRR methods are presented in Sec.~\ref{sec:result}.
The extrapolation power of the EKRR method is discussed.
The strong OES of the nuclear charge radii in Ca and Cu isotopes
and abrupt kinks across the neutrons $N=126$ and 82 shell closures were investigated.
Finally, a summary is presented in Sec.~\ref{sec:summary}.

\section{Theoretical framework}\label{sec:ekrr}
The KRR method was successfully applied to improve the descriptions of
nuclear charge radii obtained using several widely used phenomenological formulae~\cite{Ma2022_CPC46-074105}.
To include odd-even effects, the KRR function
$S(\bm{x_j})=\sum_{i=1}^m K(\bm{x_j},\bm{x_i})\alpha_i$
is extended to be the EKRR function~\cite{Wu2021_PLB819-136387}
\begin{equation}\label{eq:1}
S(\bm{x_j})=\sum_{i=1}^m K(\bm{x_j},\bm{x_i})\alpha_i +
            \sum_{i=1}^m K_\mathrm{oe}(\bm{x_j},\bm{x_i})\beta_i \ ,
\end{equation}
where $\bm{x_i}$ are the locations of the nuclei in the nuclear chart, with $\bm{x}_i=(Z_i,N_i)$.
$m$ is the number of training data points, $\alpha_i$  and $\beta_i$  are the weights,
$K(\bm{x_j},\bm{x_i})$ and $K_\mathrm{oe}(\bm{x_j},\bm{x_i})$
are kernel functions that characterize the similarity between the data.
In this study, a Gaussian kernel was adopted, which is expressed as
\begin{equation}\label{eq:2}
K(\bm{x_j},\bm{x_i})=\mathrm{exp}(-||\bm{x_i}-\bm{x_j}||^2/2\sigma^2) \ ,
\end{equation}
where $||\bm{x_i}-\bm{x_j}||=\sqrt{(Z_i-Z_j)^2+(N_i-N_j)^2}$ is the distance between two nuclei.
$K_\mathrm{oe}(\bm{x_j},\bm{x_i})$
was introduced to enhance the correlations between nuclei with the same number parity of neutrons and protons,
which can be written as:
\begin{equation}\label{eq:3}
K_\mathrm{oe}(\bm{x_j},\bm{x_i}) = \delta_\mathrm{oe}(\bm{x_j},\bm{x_i})
\mathrm{exp}(-||\bm{x_i}-\bm{x_j}||^{2}/2\sigma_\mathrm{oe}^2) \ .
\end{equation}
$\delta_\mathrm{oe}(\bm{x_j},\bm{x_i})=1$ (0) if the two nuclei have
the same (different) number parities of protons and neutrons.
$\sigma$ and $\sigma_\mathrm{oe}$ are hyperparameters
We defined the range affected by the kernel.

The kernel weights $\alpha_i$ and $\beta_i$ are determined by minimizing the following loss function:
\begin{equation}\label{eq:4}
L(\bm{\alpha},\bm{\beta})=\sum_{i=1}^m\left[S(\bm{x_i})-y(\bm{x_i})\right]^2 +
 \lambda\bm{\alpha}^T\bm{K}\bm{\alpha}+\lambda_\mathrm{oe}\bm{\beta}^T\bm{K}_\mathrm{oe}\bm{\beta} \ .
\end{equation}
The first term is the variance between the training data $y(\bm{x_i})$ and the EKRR prediction $S(\bm{x_i})$.
The second and third terms are regularizers, where the hyperparameters $\lambda$ and $\lambda_{\mathrm{oe}}$
determine the regularization strength and are adopted to reduce the risk of overfitting.

By minimizing the loss function [Eq. (\ref{eq:4})], we obtain
\begin{eqnarray}
\bm{\beta}&=&\frac{\lambda}{\lambda_{\mathrm{oe}}}\bm{\alpha} \ , \label{eq:5} \\
\bm{\alpha}&=&\left(\bm{K}+\bm{K}_{\mathrm{oe}}
\frac{\lambda}{\lambda_{\mathrm{oe}}}+\lambda \bm{I}\right)^{-1}\bm{y} \ . \label{eq:6}
\end{eqnarray}
According to Eq.~(\ref{eq:5}), the EKRR function [Eq.~(\ref{eq:1})]
can be written as a standard KRR function:
\begin{equation}\label{eq:7}
S(\bm{x_{j}})=\sum K'(\bm{x_{j}},\bm{x_{i}})\alpha_{i} \ ,
\end{equation}
where $K'(\bm{x_j},\bm{x_i})$ is the remodulation kernel.
\begin{equation}\label{eq:8}
K'(\bm{x_j},\bm{x_i})=K(\bm{x_j},\bm{x_i})+\frac{\lambda}{\lambda_\mathrm{oe}}K_\mathrm{oe}(\bm{x_j},\bm{x_i}) \ .
\end{equation}
According to Eq.~(\ref{eq:5}), the number of weight parameters
in the EKRR method is identical to that in the original KRR method.

\section{Numerical details}\label{sec:details}
In this study, 1014 experimental data points with $Z \geq 8$
were considered and obtained from Refs.~\cite{Angeli2013_ADNDT99-69, Li2021_ADNDT140-101440}.
The EKRR function~(\ref{eq:7}) was trained to reconstruct the residual radius:
i.e., the deviations $\Delta R(N,Z) = R^{\rm{exp}}(N,Z) -R^{\rm{th}}(N,Z)$
between the experimental data $R^{\rm{exp}}(N,Z)$ and the predictions $R^{\rm{th}}(N,Z)$
for the following five nuclear models.
\begin{enumerate}[(i)]
\item  The widely used phenomenological formula $R_c=r_A \left[1-b(N-Z)/A\right]A^{1/3}$~\cite{Nerlo-Pomorska1993_ZPA344-359} with
    the parameter $r_A$=1.282 fm and $b=0.342$ was fitted by experimental data (further denoted by $A^{1/3}$).
 \item  The relativistic continuum Hartree-Bogoliubov (RCHB) theory~\cite{Xia2018_ADNDT121-122-1}.
 \item  The Hartree-Fock-Bogoliubov (HFB) model HFB25~\cite{Goriely2013_PRC88-024308}.
 \item  The Weizs\"acker-Skyrme (WS) model WS$^\ast$~\cite{Li2021_ADNDT140-101440}.
 \item  The HFB25$^\ast$ model~\cite{Li2021_ADNDT140-101440}.
\end{enumerate}

Note that by considering the nuclear shell corrections and deformations
obtained from the WS and HFB25 models,
a five-parameter nuclear charge radii formula was proposed in Ref.~\cite{Li2021_ADNDT140-101440}.
In this study, these methods are denoted as WS$^\ast$ and HFB25$^\ast$, respectively.
The parameters in the formulae of these two models were obtained from Refs.~\cite{Li2021_ADNDT140-101440}.
The RMS deviations between the experimental data and the five models ($\Delta_{\rm rms}$) are listed in Table~\ref{t:1}.
Once the weights $\alpha_i$ were obtained, the EKRR function $S(N, Z)$ was obtained for each nucleus.
Therefore, the predicted charge radius for a nucleus with neutron number $N$
and the proton number $Z$ is given by $R^{\rm EKRR} = R^{\rm th}(N, Z)+S(N,Z)$.
In this study, the KRR method was adopted for predicting charge radii for comparison.

Leave-one-out cross-validation was adopted to determine the two hyperparameters ($\sigma$ and $\lambda$)
in the KRR method and the four hyperparameters ($\sigma$, $\lambda$, $\sigma_{\rm{oe}}$ and $\lambda_{\rm{oe}}$)
in the EKRR method.
The predicted radius for each of the 1014 nuclei can be given by the
KRR/EKRR method trained on all other 1013 nuclei with a given set of hyperparameters.
The optimized hyperparameters (see Table~\ref{t:1}) are obtained when the RMS deviation between the experimental and calculated
radii reach a minimum value.

\section{Results and Discussion}\label{sec:result}

\begin{table*}[!]
\caption{\label{t:1}
The hyperparameters ($\sigma$, $\lambda$, $\sigma_{\rm oe}$ and $\lambda_{\rm oe}$)
in the KRR and EKRR method, and the RMS deviations between the experimental data
and the predictions by five different models.
The RMS deviations with (without) KRR and EKRR methods are denoted by
$\Delta_{\rm rms}^{\rm KRR}$ and $\Delta_{\rm rms}^{\rm EKRR}$ ($\Delta_{\rm rms}$). }
\begin{tabular*}{0.7\textwidth}{@{\extracolsep{\fill}}cccccccc}
\toprule[1pt]\specialrule{0em}{1pt}{1pt}
Model          &$\sigma$ & $\lambda$ & $\sigma_{\rm oe}$     & $\lambda_{\rm oe}$  &
$\Delta_{\rm rms}$~(fm)  & $\Delta_{\rm rms}^{\rm KRR}$~(fm) & $\Delta_{\rm rms}^{\rm EKRR}$~(fm)   \\
\hline
$A^{1/3}$      &  -   &  -   &  -   &  -   & 0.0672                                &   -    &    -   \\
~              & 2.84 & 0.01 &  -   &  -   &    -                                  & 0.0158 &    -   \\
~              & 2.32 & 0.01 & 2.88 & 0.02 &    -                                  &   -    & 0.0100 \\
RCHB           &  -   &  -   &  -   &  -   & 0.0350~\cite{Xia2018_ADNDT121-122-1}  &   -    &    -   \\
~              & 2.68 & 0.02 &  -   &  -   &    -                                  & 0.0157 &    -   \\
~              & 1.83 & 0.01 & 2.73 & 0.02 &    -                                  &   -    & 0.0092 \\
HFB25          &  -   &  -   &  -   &  -   & 0.0256~\cite{Goriely2013_PRC88-024308}&   -    &    -   \\
~              & 1.77 & 0.34 &  -   &  -   &    -                                  & 0.0177 &    -   \\
~              & 1.48 & 0.08 & 2.20 & 0.22 &    -                                  &   -    & 0.0130 \\
WS$^{\ast}$    &  -   &  -   &  -   &  -   & 0.0210~\cite{Li2021_ADNDT140-101440}  &   -    &    -   \\
~              & 0.70 & 0.01 &  -   &  -   &    -                                  & 0.0155 &    -   \\
~              & 1.54 & 0.02 & 2.46 & 0.03 &    -                                  &   -    & 0.0096 \\
HFB25$^{\ast}$ &  -   &  -   &  -   &  -   & 0.0254~\cite{Li2021_ADNDT140-101440}  &   -    &   -    \\
~              & 0.68 & 0.01 &  -   &  -   &    -                                  & 0.0182 &    -   \\
~              & 1.35 & 0.05 & 2.21 & 0.08 &    -                                  &   -    & 0.0120 \\
\bottomrule[1pt]
\end{tabular*}
\end{table*}

\begin{figure}[!]
\centering
\includegraphics[width=0.8\columnwidth]{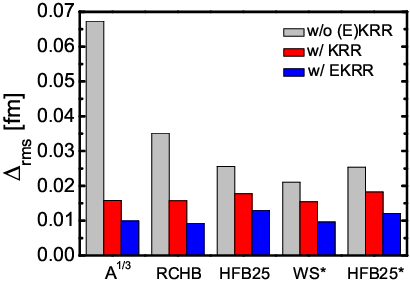}
\caption{(Color online) The RMS deviations between the experimental data and the predictions of five different models with and
without the KRR/EKRR method.}
\label{fig1:rms}
\end{figure}

Table~\ref{t:1} lists the hyperparameters ($\sigma$, $\lambda$) in the KRR method and
($\sigma$, $\lambda$, $\sigma_{\rm oe}$ and $\lambda_{\rm oe}$) using the EKRR method as well as the RMS deviations between the
experimental data and the predictions of the five models.
The RMS deviations with (without) KRR and EKRR are denoted by
$\Delta_{\rm rms}^{\rm KRR}$ and $\Delta_{\rm rms}^{\rm EKRR}$ ($\Delta_{\rm rms}$).
With the exception of the phenomenological $A^{1/3}$-formula
all other models provided a good global description of the nuclear charge radii,
especially for the WS$^{\ast}$.
It should be noted that a spherical shape is considered in the RCHB theory when investigating the entire nuclear
landscape~\cite{Xia2018_ADNDT121-122-1}.
Therefore, its RMS deviation is slightly larger than that for the nonrelativistic model HFB25.
To date, only even-even nuclei have been calculated in the deformed relativistic
Hartree-Bogoliubov theory on a continuum (DRHBc)~\cite{Zhang2020_PRC102-024314, Zhang2022_ADNDT144-101488}.
The description of the nuclear charge radii can be further improved when
all nuclei in the nuclear chart are calculated using this model.
It can also be observed that HFB25 and HFB25$^{\ast}$ yield similar RMS deviations when describing the nuclear charge radii.
After the KRR method had been considered, all RMS deviations for these five models
could be significantly reduced to approximately 0.015-0.018~fm, particularly for the $A^{1/3}$ formula.
Interestingly, the RMS deviations of the HFB25 and HFB25$^\ast$ models
were smaller than those of  the $A^{1/3}$ formula and the RCHB model without the KRR method.
However, after the KRR method was considered, the situation was reversed.
After considering the odd-even effects, the predictive powers of the five models
were further improved by the EKRR method compared with the KRR method.
The RMS deviation was further reduced by approximately 0.006 fm for the five models,
with the exception of the HFB25 model, for which it was reduced to less than 0.005~fm.
The RMS deviations of the three models ($A^{1/3}$ formula, RCHB and WS$^{\ast}$) were less than 0.01~fm,
whereby the smallest was for the RCHB model with an RMS deviation equal to 0.0092~fm.
This is the best result for nuclear charge radii predictions
using the machine learning approach, as far as we are aware.
Here, we show the typical RMS deviations of some popular machine learning approaches.
\begin{enumerate}[(i)]
\item artificial neural network: 0.028 fm~\cite{Wu2020_PRC102-054323};
\item Bayesian neural network: 0.014 fm~\cite{Dong2023_PLB838-137726};
\item radial basis function approach: 0.017 fm~\cite{Li2023_NPR_40-31}.
\end{enumerate}
Note that if the full nuclear landscape is calculated using the DRHBc theory,
the description of the nuclear charge radii can still be improved using the EKRR method.
To show Table~\ref{t:1} in a more visual manner, a comparison of these five models is also shown in Fig.~\ref{fig1:rms}.

\begin{figure}[h]
\centering
\includegraphics[width=1.0\columnwidth]{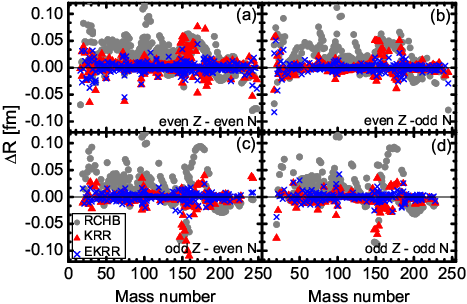}
\caption{(Color online) Radius differences $\Delta R$ between the experimental data and the calculations of
the RCHB model (grey solid circles), the KRR method (red triangles), and the EKRR method (blue crosses)
for (a) even-even, (b) even-odd, (c) odd-even, and (d) odd-odd nuclei.}
\label{fig2:deltR}
\end{figure}

Figure~\ref{fig2:deltR} shows the differences in the radii between the experimental data
and the calculations of the RCHB model (grey solid circles), KRR method (red triangles)
and the EKRR methods (blue crosses).
Because the improvements achieved by the KRR and EKRR methods for the five models mentioned above were similar,
we consider only the RCHB model as an example.
In order to study the odd-even effects included in the EKRR method,
the data were divided into four groups characterized by even or odd proton numbers $Z$ and neutron numbers $N$,
that is, even-even, even-odd, odd-even, and odd-odd.
Clearly, the predictive power of the RCHB model could be further improved by
using the EKRR method compared with the original KRR method.
The significant improvement of the EKRR method is mainly due to the consideration of the odd-even effects,
which eliminates the staggering behavior of radius deviations owing to
the odd and even numbers of nucleons using the KRR method.
It can be seen that when the mass number is $A \sim 150$, the predictions of the KRR method
exhibit significant deviations from the data,
which can be significantly improved using the EKRR method.
This is clear evidence of the importance of considering the odd-even effects in predictions of the nuclear charge radius.

\begin{figure*}[!]
\centering
\includegraphics[width=0.8\textwidth]{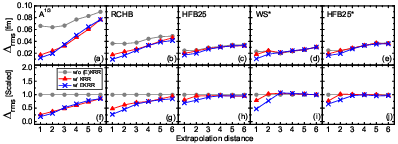}
\caption{(Color online) Comparison of the extrapolation ability of the KRR and EKRR methods for the neutron-rich region
by considering six test sets with different extrapolation distances.
The upper panels (a)-(e) show the RMS deviations of the KRR and EKRR methods.
The lower panels (f)-(j) show the RMS deviations scaled to the corresponding RMS deviations for
these five models without KRR/EKRR corrections.}
\label{fig3:extrap}
\end{figure*}

To investigate the extrapolation abilities of the KRR and EKRR methods for neutron-rich regions,
the 1014 nuclei with known charge radii were redivided into one training set and six test sets as follows:
For each isotopic chain with more than nine nuclei,
the six most neutron-rich nuclei were selected and classified into six test sets
based on the distance from the previous nucleus.
Test set 1 (6) had the shortest (longest) extrapolation distance.
This type of classification is the same as that used in our previous study~\cite{Ma2022_CPC46-074105}.
The hyperparameters obtained by leave-one-out cross-validation in
the KRR/RKRR method remained the same in the following calculations:

RMS deviations of the KRR and EKRR methods for different extrapolation steps
for the five models are shown in Figs.~\ref{fig3:extrap}(a)-(e).
A clearer comparison of the RMS deviations scaled to the corresponding RMS deviations of
the five models without KRR/EKRR corrections are shown in Figs.~\ref{fig3:extrap}(f) and (j).
Regardless of whether the KRR or EKRR method is considered,
the RMS deviation increased with the extrapolation distance.
For the $A^{1/3}$ formula and the RCHB model, the KRR/EKRR method could improve
the radius description for all extrapolation distances.
For the other three models, the KRR method only improved the radius description for an extrapolation distance of one or two, which
could be further improved after considering the odd-even effects with the EKRR method.
This indicates that the KRR/EKRR method loses its extrapolation power
at extrapolation distances larger than 3 for these three models.
This is due to the charge radii calculated using these three models, which
were quite good, and their RMS deviations, which were already sufficiently small.
The KRR/EKRR method automatically identifies the extrapolation distance limit
owing to the hyperparameters $\sigma$ and $\sigma_{\rm oe}$ being optimized using the training data.
Refs.~\cite{Wu2020_PRC101-051301, Wu2021_PLB819-136387} demonstrated that the KRR and EKRR methods
lose their predictive power at larger extrapolation distances (approximately six),
when predicting the nuclear mass using the mass model WS4~\cite{Wang2014_PLB734-215}.
This may be due to much more mass data existed than the charge radii,
and the KRR/EKRR networks can be trained better with more data.
In general, the EKRR method has a better predictive power
than the KRR method for an extrapolation distance of less than 3.
For an extrapolation distance greater than 3,
the results of the KRR and EKRR methods were similar in most cases.
Almost none of these extrapolations exhibited overfitting,
except for WS$^{\ast}$ at an extrapolation distance of 3,
and this overfitting was quite small.
This indicates that both the KRR and EKRR methods have good extrapolation powers and can avoid the risk of
overfitting to a large extent.

\begin{figure}[!]
\centering
\includegraphics[width=0.95\columnwidth]{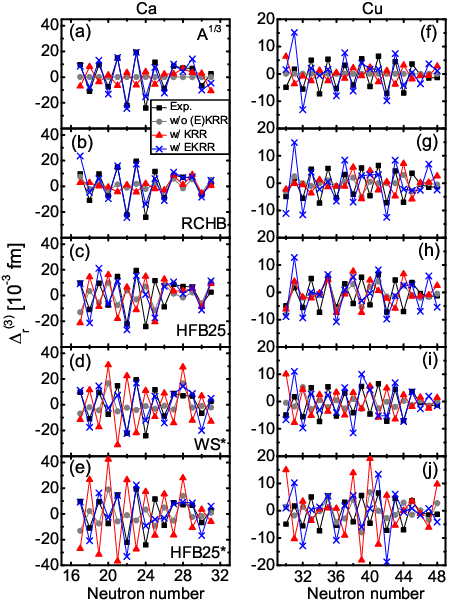}
\caption{(Color online) Comparison of experimental and calculated OES of the charge radii ($\Delta^{(3)}_r$)
of the calcium (left panels) and copper (right panels) isotopes.
The experimental data are shown as black squares.
The calculation results of these five models are shown as grey solid circles,
and the calculation results of the KRR and EKRR models
are shown as red triangles and blue crosses, respectively.}
\label{fig4:oes}
\end{figure}

The observation of the strong OES of the charge radii throughout the nuclear
landscape provides a particularly stringent test for nuclear theory.
To examine the predictive power of the EKRR method,
which is improved by considering the odd-even effects compared with the original KRR method,
in the following we will investigate the recently observed OES of the radii in calcium and
copper isotopes~\cite{Ruiz2016_NatPhys12-594, Miller2019_NatPhys15-432, Groote2020_NatPhys16-620}.
Similar to the gap parameter, the OES parameter for the charge radii is defined as:
\begin{equation}\label{eq:9}
\Delta_r^{(3)}(Z,N)=\frac{1}{2}[r(Z,N-1) - 2r(Z,N) + r(Z,N+1)] \ ,
\end{equation}
where $r(Z, N)$ is the RMS charge radius of a nucleus with proton number $Z$ and neutron number $N$.

Figure~\ref{fig4:oes} compares the experimental and calculated OES results for
radii ($\Delta^{(3)}_r$) of the calcium (left panels) and copper (right panels) isotopes.
The experimental data show that for the calcium isotopes [Figs.~\ref{fig4:oes}(a)-(e)]
strong OES exists between $N=20$ and $ 28 $ and that a reduction in the OES appears for $N \ge 28$.
Only RCHB theory could reproduce the trend of the experimental OES without KRR/EKRR corrections.
However, the amplitude of the calculated OES was significantly less pronounced than that of the experimental data.
Interestingly, after considering the KRR corrections, the calculated OES worsened for $N < 28$, particularly when the phase of the
OES was opposite to that of the data.
The $A^{1/3}$-formula had no OES over the entire isotopic chain
and the WS$^{\ast}$ model has a weak OES except at the $N=20$ and 28 shell closures.
The OES in the HFB25 and HFB25$^{\ast}$ models were slightly higher.
However, they were still weak compared with the data.
Note that although OES can be obtained in the WS$^{\ast}$, HFB25 and HFB25$^{\ast}$ models, the phases of the calculated OES are
opposite to those of the experimental data.
Considering the KRR method, the OES in these four models increased,
particularly for the WS$^{\ast}$ and HFB25$^{\ast}$ models
for which the calculated OES were stronger than those of the data.
However, the OES in these models were still opposite to those in the data.
Therefore, although the KRR method improves the description of the charge radius to a large extent, it was difficult to reproduce
the observed OES.
After considering the EKRR method, the experimental OES values could be reproduced quite well, especially for the $A^{1/3}$ formula
and RCHB theory for copper isotopes [Figs.~\ref{fig4:oes}(f)-(j)].
This situation is similar to that of the calcium isotopes.
However, the description of Cu isotopes is not as accurate as that of Ca isotopes when considering the EKRR corrections.
The OES is overestimated in all these calculations for $N < 33$ and $N > 46$.
In addition, the phases of the OES between $N=38$-40 were not well reproduced.
However, the EKRR approach can improve the description of OES to a
large extent compared with the original theory.
This indicates that after considering the odd-even effects,
shell structures and many-body correlations, which are important for OES,
can be learned well using an EKRR network.

\begin{figure}[h]
\centering
\includegraphics[width=0.95\columnwidth]{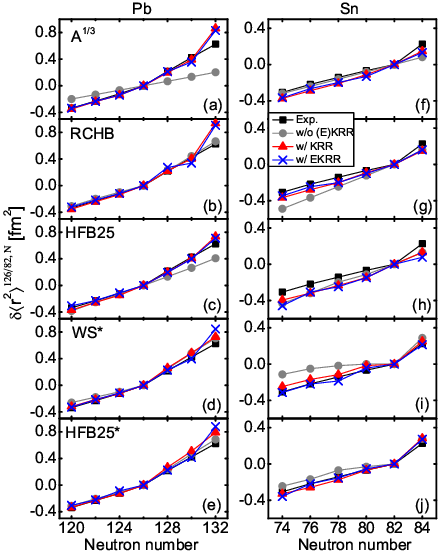}
\caption{(Color online) Comparison of experimental and calculated differential mean-square charge radius
$\delta \langle r^2 \rangle^{N',N} = \langle r^2 \rangle^N - \langle r^2 \rangle^{N'}$
for some even-even (a)-(e) Pb (relative to $^{208}$Pb, $N'=126$)
and (f)-(j) Sn (relative to $^{132}$Sn, $N'=82$) isotopes.
The experimental data are shown as black squares.
The results of these five models are shown as grey solid circles,
and the calculation results of the KRR and EKRR models
are shown as red triangles and blue crosses, respectively.}
\label{fig5:kinks}
\end{figure}

Similar to OES, abrupt kinks across the neutron shell closures
provide a particularly stringent test for nuclear theory.
In the present study, Pb and Sn isotopes were considered as examples for investigating the kinks across neutrons with $N=126$ and 82
shell closures.
Figure~\ref{fig5:kinks} compares the experimental and calculated differential mean-square charge radii.
$\delta \langle r^2 \rangle^{N',N} = \langle r^2 \rangle^N -\langle r^2 \rangle^{N'}$
for some, and even for Pb [Figs.~\ref{fig5:kinks}(a)-(e)] (relative to $^{208}$Pb, $N'=126$)
and Sn [Figs.~\ref{fig5:kinks}(f)-(j)] (relative to $^{132}$Sn, $N'=82$).
It can be observed that for Pb isotopes
the RCHB theory can reproduce the kink at $N=126$ perfectly [Fig.~\ref{fig5:kinks}(b)].
In the $A^{1/3}$ formula and HFB25 model, there is no kink [Figs.~\ref{fig5:kinks}(a) and (c)].
The kink could be reproduced using the WS$^{\ast}$ and HFB25$^{\ast}$ models,
but with a slight overestimation [Figs.~\ref{fig5:kinks}(d) and (e)].
The results obtained by considering the KRR and EKRR methods were similar.
There are several interpretations of kinks~\cite{Sharma1995_PRL74-3744, Perera2021_PRC104-064313,
Nakada2015_PRC91-021302, Nakada2015_PRC92-044307, Fayans2000_NPA676-49}.
Our results indicate that kinks may not be connected to  odd-even effects, such as pairing correlations.
The well-reproduced kinks also provide a test of the proposed KRR/EKRR method.
The kinks at $N=126$ in all five models could be reproduced quite well,
but the calculated differential mean-square charge radius at $N=132$ was too large compared with the data.
For the Sn isotopes, only the WS$^{\ast}$ and HFB25$^{\ast}$ models reproduced the kink at $N=82$.
However, the absolute values of the calculated $\delta \langle r^2 \rangle$
from $N=74$-78 are small compared with the data, especially for the WS$^{\ast}$ model.
After applying the KRR/EKRR method, the results reproduced the data quite well.
It also can be seen that the KRR/EKRR corrections to the $A^{1/3}$ formula and HFB25 model
are inconspicuous. Therefore, the kink at $N=82$ cannot be reproduced using the KRR/EKRR method.
For the RCHB model, the differential mean-square charge radii calculated from $N=74$ to-80 were
improved, and a kink appeared, but was still slightly weaker compared with the data.

\section{Summary}\label{sec:summary}

In summary, the extended kernel ridge regression method with odd-even effects
was adopted to improve the description of the nuclear charge radius by using five commonly used nuclear models.
The hyperparameters of the KRR and EKRR methods for each model were determined using leave-one-out cross-validation.
For each model, the resultant root-mean-square deviations of the 1014 nuclei
with proton number $Z \geq 8$ could be significantly reduced
to 0.009-0.013~fm after considering a modification with the EKRR method.
The best among them was the RCHB model, with a root-mean-square deviation of 0.0092~fm,
which is the best result for nuclear charge radii predictions using the machine learning approach as far as we know.
The extrapolation abilities of the KRR and EKRR methods for the neutron-rich region were
examined and it was found that after considering odd-even effects,
the extrapolation power could be improved compared with that of the original KRR method.
Strong odd-even staggering of nuclear charge radii in Ca and Cu isotopes
was investigated and reproduced quite well using the EKRR method.
This indicates that after considering the odd-even effects,
shell structures and many-body correlations can be learned quite well using the EKRR network.
Abrupt kinks across the neutron $N=126$ and 82 shell closures were also investigated.


%

\end{document}